# THE LEPTO-VARIANCE OF STOCK RETURNS

*Vassilis Polimenis*
Aristotle University
polimenis@auth.gr

### ABSTRACT

The Regression Tree (RT) sorts the samples using a specific feature and finds the split point that produces the maximum variance reduction from a node to its children. Our key observation is that the best factor to use (in terms of MSE drop) is always the target itself, as this most clearly separates the target. Thus, using the target as the splitting factor provides an upper bound on MSE drop (or lower bound on the residual children MSE). Based on this observation, we define the k-bit lepto-variance $\lambda k^2$ of a target variable (or equivalently the lepto-variance at a specific depth k) as the variance that cannot be removed by any regression tree of a depth equal to k. As the upper bound performance for any feature, we believe $\lambda k^2$ to be an interesting statistical concept related to the underlying structure of the sample as it quantifies the resolving power of the RT for the sample. The max variance that may be explained using RTs of depth up to k is called the sample k-bit macro-variance. At any depth, total sample variance is thus decomposed into lepto-variance $\lambda^2$ and macro-variance $\mu^2$. We demonstrate the concept, by performing 1- and 2-bit RT based lepto-structure analysis for daily IBM stock returns.

## 1. INTRODUCTION

In many machine learning applications, from biostatistics to finance there is an interest in finding features that explain a large fraction of the total variance of a target outcome. For example, in financial analysis, financial factors are used as features and there is an interest in finding the factors that explain a large fraction of the total stock price variability. Stock return variance that cannot be explained by broad market financial factors is considered idiosyncratic for the specific stock. In regression analysis, the residual sum of squares (RSS) or the idiosyncratic variance depend on the factors used in the regression. In general, we can get lower residual variance by adding an extra factor. Thus, idiosyncratic or unexplained variance depends on the factors used to explain the dependent variable. Here, by using the Regression Tree (RT) methodology, we define a new type of idiosyncratic variance, the **lepto-variance** that is uniquely defined for a sample regardless of the set of factors used to explain the variable. We define the **macro-variance** of a sample as the upper bound of the sample variability that may be explained by any attribute. Lepto- and macro-variance are features that depend solely on sample structure. Similar to the resolving power of an optical system (such as a microscope or telescope) that places a limit on its ability to distinguish images or reproduce fine detail, the lepto-variance of a target places a limit on the ability of a k-bit RT in explaining fine structure in the sample.

## 2. THE RESOLVING POWER OF A REGRESSION TREE

Usually, Regression Trees (RT) provide a binary splitting of the sample space with minimization criterion the Residual Sum of Squares $\Sigma_j (y_j - \bar{y}_j)^2$, with $\bar{y}_j$ the average y for the subspace that includes $y_j$. Finding the best binary partition of the feature space for a regression tree is computationally infeasible. Hence, a recursive greedy algorithm is used. When splitting a categorical predictor that takes q possible unordered values, $2^{q-1} - 1$ possible partitions of the q values into two groups need to be evaluated. Such computation becomes prohibitive for large q. There are various "concavity" theorems that simplify the problem (see [Ripley (1996)] and [Hastie et al. (2009)]). Even before the publication of the first RT algorithms, [Fisher (1958)] showed that for a continuous-valued target Y the least squares partition of a set is contiguous. [Breiman et al. (1984)] extended for a decision tree with binary (2-class) target Y.

Instead of performing a search in an exponential number of possible binary partitions, a simplified linear search of only q-1 partitions where attribute values are sorted based on their strength of correlation with the target outcome suffices. Starting at the root node of the tree (#0) which includes all samples, the algorithm will consider all possible splitting factors (x) and split points (c) that define a pair of half-planes. For each factor x, the best split point c can be very quickly determined and hence by scanning through all factors, the best pair (x, c) can be feasibly determined, and the algorithm recursively repeats the splitting process for each of the two children nodes.

Effectively, at each node #j the RT sorts the sub-sample of this node $S_j$ using the chosen split factor $x_j$ and finds the split point $c_j$ that produces the maximum MSE drop from the node to its children $L_j$ and $R_j$. We may think of this MSE drop (variance[i] reduction) as an (informational or impurity) gain from splitting via this factor. With no loss of generality, assume that the left child L contains the small y values (i.e. assume $\bar{y}_L$ = mean(y | in L sub-sample) < $\bar{y}_R$ = mean(y | in R sub-sample)). Generally, sorting a sample based on a factor $x_j$ will not produce a sorted target y.

**Definition 1**. A binary split of S into L and R is *sorted* if all target values y in L are smaller than all target values in R.

Equivalently, a split is sorted if the maximum target value in L is smaller than the minimum target value in R.

We provide an extension of the [Fisher (1958)] theorem on grouping, with the following lemma for Regression Trees.

**Lemma 1.** In a Regression Tree, using the target itself as the predictor provides an upper bound on the MSE reduction.

**Proof.** We first show that any unsorted split is strictly dominated by a sorted split. Let us assume an unsorted split of a sample S into L and R (with no loss of generality, assume $\bar{y}_L < \bar{y}_R$). Then the maximum value of the left sub-tree u1 is larger than the minimum value of the right sub-tree u0

$$u1 = \max(L) > u0 = \min(R)$$

But then, we can get a better split by swapping u0 with u1 into L and R respectively. Because moving u0 into L and u1 into R, will move the center of the L sub-sample farther to the left and the center of the right sub-sample farther to the right, thus producing a larger separation between the left and right sub-samples without changing the relative sample sizes. By the law of total variance, a larger between-group variability means a smaller within-group variability and thus a better split.

Since for any unsorted split there exists a strictly better sorted split, the best binary split is always a sorted split. Using the target on itself would most clearly separate the target, as it can generate all sorted splits. Thus, no predictor can ever perform better (in terms of MSE drop) than the target itself.

## 2.1 The 1-bit Lepto-structure of a sample

To clarify the ideas an example is presented below. On Table 1 we see the values for two financial factors f1 and f2 and a hypothetical stock y for 8 days. There is an outlier return for the stock at t=3.

*Table 1.* Hypothetical returns of two factors f1 and f2 and a stock y for 8 days (in percentages)

| t | f1 | f2 | y | |
|---|-----|-----|------|-----------|
| 1 | 2.0 | 2.0 | 1.5 | |
| 2 | 1.8 | 6.2 | -1.0 | |
| 3 | 5.0 | 1.8 | 4.0 | * outlier |
| 4 | 7.0 | 4.0 | 2.0 | |
| 5 | 6.0 | 6.0 | 1.0 | |
| 6 | 4.8 | 5.8 | -0.5 | |
| 7 | 2.2 | 5.0 | -2.0 | |
| 8 | 1.0 | 1.0 | 0.0 | |

When returns are simply sorted by the date they occur, the best split is a group of the first 5 samples. Table 2a depicts MSE for the left and right children for all possible splits from position 1 to 8 and the total weighted children MSE in the last column.[ii] Figure 1a shows the optimal resulting depth 1 tree.

In this case the weighted MSE drops from 3.172 to 1.896 with an MSE drop of 40.23% = 100% - 1.896 / 3.172. In terms of outcome values, the split is {-0.5, -2.0, 0.0} and {1.5, -1.0, 4.0, 2.0, 1.0} - an unsorted split strictly inferior to the one we would get by swapping 0.0 with -1.0.

When the 1st factor f1 is used (see Table 2b), the split point is f1<4.9 (i.e. the middle point between 4.8 and 5.0). The 5 samples t=1, 2, 6, 7, 8 belong to the left branch, while t=3, 4, 5 are put in the right branch. In this case the minimum weighted children MSE becomes 1.421 with a 55.21% MSE drop. Figure 1b shows the optimal RT. When the

2nd factor f2 is used, the split point is f2 < 4.5 (i.e. the middle point between 4.0 and 5.0). The 4 samples t=1, 3, 4, 8 belong to the left branch, while t=2, 5, 6, 7 are put in the right branch. In this case the weighted MSE becomes 1.609 with a 49.26% MSE drop. Figure 1c shows the optimal RT.

As proven in Lemma 1, regressing a series on itself achieves the upper bound of MSE drop and the lowest children MSE. In particular, here when the factor used is y, the optimal split point is $y \leq 0$ (i.e. equivalently y < .5, the middle point between 0 and 1.0). The 4 samples t=2, 6, 7, 8 belong to the left branch, while t=1, 3, 4, 5 are put in the right branch (see Table 2d). In this case, the lowest possible MSE becomes $\lambda 1^2 = .922$ with a 70.94% MSE drop. The symbol $\lambda 1^2$ is used for depth 1 **lepto-variance**. For obvious reasons, $\lambda 1^2$ is also called the *1-bit lepto-variance* of the sample.

**Figure 1a.** *Split for the optimal depth=1 RT when simple time (t) is used as a factor at t<5.5 (5.5 is the mid-point between t=5 and 6)*

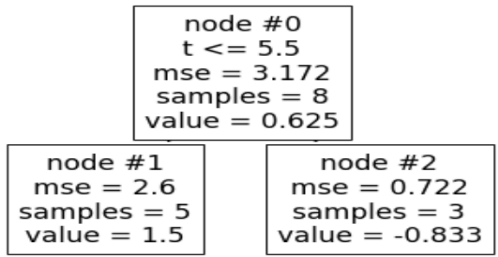

**Table 2a.** *Optimal depth=1 RT when simple time (t) is used as the splitting factor; i.e. samples are sorted based on their t values.*

| t | f1 | f2 | y | mse L | mse R | L+R |
|---|-----|-----|------|-------|-------|-------|
| 1 | 2.0 | 2.0 | 1.5  | 0.000 | 3.500 | 3.063 |
| 2 | 1.8 | 6.2 | -1.0 | 1.563 | 3.646 | 3.125 |
| 3 | 5.0 | 1.8 | 4.0  | 4.167 | 1.840 | 2.713 |
| 4 | 7.0 | 4.0 | 2.0  | 3.172 | 1.172 | 2.172 |
| 5 | 6.0 | 6.0 | 1.0  | 2.600 | 0.722 | 1.896 |
| 6 | 4.8 | 5.8 | -0.5 | 2.722 | 1.000 | 2.292 |
| 7 | 2.2 | 5.0 | -2.0 | 3.561 | 0.000 | 3.116 |
| 8 | 1.0 | 1.0 | 0.0  | 3.172 |       |       |
|   |     |     |      | MSE   |       | 1.896 |
|   |     |     |      | info gain |   | 40.23% |

**Figure 1b.** *Split for the optimal depth=1 RT when the first factor is used at f1<4.9 (4.9 is used as the mid-value between 4.8 and 5.0).*

© Vassilis Polimenis 2022

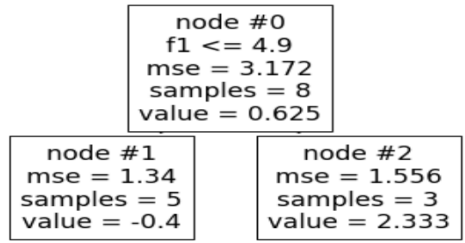

**Table 2b.** Optimal depth=1 RT when the first factor f1 is used; i.e. samples are sorted based on their f1 values.

| t | f1 | f2 | y | mse L | mse R | L+R |
|---|----|----|----|-------|-------|------|
| 8 | 1.0 | 1.0 | 0.0 | 0.000 | 3.561 | 3.116 |
| 2 | 1.8 | 6.2 | -1.0 | 0.250 | 3.583 | 2.750 |
| 1 | 2.0 | 2.0 | 1.5 | 1.056 | 4.240 | 3.046 |
| 7 | 2.2 | 5.0 | -2.0 | 1.672 | 2.672 | 2.172 |
| 6 | 4.8 | 5.8 | -0.5 | 1.340 | 1.556 | 1.421 |
| 3 | 5.0 | 1.8 | 4.0 | 3.806 | 0.250 | 2.917 |
| 5 | 6.0 | 6.0 | 1.0 | 3.316 | 0.000 | 2.902 |
| 4 | 7.0 | 4.0 | 2.0 | 3.172 | | |
| | | | | MSE | 1.421 | |
| | | | | info gain | 55.21% | |

**Figure 1c.** Split for the optimal depth=1 RT when the second factor is used at f2<4.5

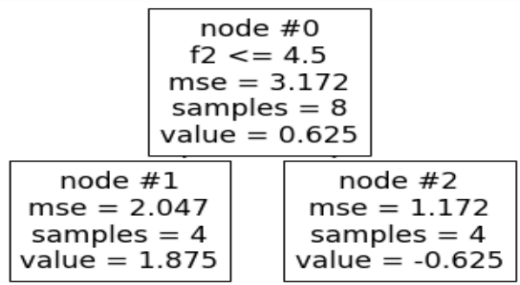

**Table 2c.** *Optimal depth=1 RT when the 2nd factor is used; i.e. samples are sorted based on their f2 values.*

| t | f1 | f2 | y | mse L | mse R | L+R |
|---|-----|-----|------|-------|-------|-------|
| 8 | 1.0 | 1.0 | 0.0 | 0.000 | 3.561 | 3.116 |
| 3 | 5.0 | 1.8 | 4.0 | 4.000 | 2.056 | 2.542 |
| 1 | 2.0 | 2.0 | 1.5 | 2.722 | 2.040 | 2.296 |
| 4 | 7.0 | 4.0 | 2.0 | 2.047 | 1.172 | 1.609 |
| 7 | 2.2 | 5.0 | -2.0 | 4.040 | 0.722 | 2.796 |
| 6 | 4.8 | 5.8 | -0.5 | 3.722 | 1.000 | 3.042 |
| 5 | 6.0 | 6.0 | 1.0 | 3.194 | 0.000 | 2.795 |
| 2 | 1.8 | 6.2 | -1.0 | 3.172 | | |
| | | | | | MSE | 1.609 |
| | | | | | info gain | 49.26% |

**Definition 2.** The 1-bit sample lepto-variance $\lambda 1^2$ of a target variable is defined as the MSE of a depth 1 RT of the target on itself.

Effectively one may think of the remaining variance fraction 29.06% as structure beyond the resolving power of depth 1 trees; no factor may ever explain more than 70.94% of the y variability with a binary depth 1 RT. Thus, the ability of a factor to explain the target via a depth 1 RT should be compared against the benchmark of the ability of the target to explain itself (i.e. 70.94% in this case). The remaining 29.06% is variance due to the **lepto-structure** of the target (the terms lepto-variance and lepto-structure are used interchangeably). Figure 1d shows the optimal resulting tree.

**Figure 1d.** *Split for the optimal depth=1 RT when the target outcome is also used as the predictor (i.e. factor is y).*

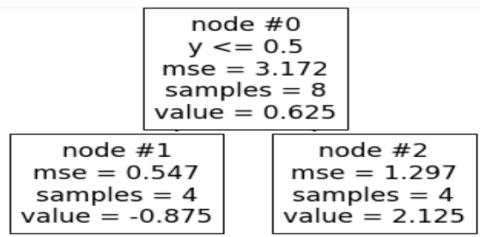

**Table 2d.** *Optimal depth=1 RT when the stock return vector is regressed on itself (i.e. factor is y); i.e. samples are sorted based on their y values.*

| t | f1 | f2 | y | mse L | mse R | L+R |
|---|-----|-----|------|-------|-------|-------|
| 7 | 2.2 | 5.0 | -2.0 | 0.000 | 2.500 | 2.188 |
| 2 | 1.8 | 6.2 | -1.0 | 0.250 | 2.139 | 1.667 |
| 6 | 4.8 | 5.8 | -0.5 | 0.389 | 1.760 | 1.246 |
| 8 | 1.0 | 1.0 | 0.0 | 0.547 | 1.297 | 0.922 |
| 5 | 6.0 | 6.0 | 1.0 | 1.000 | 1.167 | 1.063 |
| 1 | 2.0 | 2.0 | 1.5 | 1.389 | 1.000 | 1.292 |
| 4 | 7.0 | 4.0 | 2.0 | 1.765 | 0.000 | 1.545 |
| 3 | 5.0 | 1.8 | 4.0 | 3.172 | | |
| | | | | | MSE | 0.922 |
| | | | | | info gain | 70.94% |

## 2.2 The 2-bit Lepto-structure of a sample

The concept of lepto-variance of a sample may also be defined for trees of a maximum depth larger than 1. As we move deeper down on a RT there will always be less residual variance. The argument of Lemma 1 will still be valid; at any node, the best split is always a sorted split. But the greediness of the RT may generate a situation where sorting in a split is sub-optimal (by not using the entire allowed max depth in some branches). Thus, although highly unlikely for realistic samples and relatively small depths, utilizing the target itself as the splitting variable may not always achieve the lowest residual error for an *allowed max depth*. For example, sorting the initial sample {0,-2,4,1} will isolate 4 prematurely. Thus, the RT will not be allowed to grow a full 2-bit tree structure that would explain the entire variability. But it will still achieve the lowest RSS at any *average* depth (1.75 bits for the example).[iii]

**Definition 3.** For any depth k, the k-bit sample lepto-variance $\lambda_k^2$ of a target variable is defined as the MSE of an (average) depth k RT of the target on itself.

The notation $\mu_1^2$ is used to denote the max variance drop (**1-bit macro-variance**) on a root node; i.e. the MSE drop when the target vector is regressed on itself. Similarly, we use $\mu_2^2$ to denote the max variance drop when using a 2-bit (depth 2) RT instead of a 1-bit RT etc. Total sample variance then equals

$$\sigma^2 = \lambda_0^2 = \mu_1^2 + \lambda_1^2 = \mu_2^2 + \lambda_2^2 = ... = \mu_j^2 + \lambda_j^2 \qquad (1)$$

Thus, the k-bit lepto-variance is the minimum residual variance of any RT with depth k. The concept is made clear for 2-bit RTs with the hypothetical 8 sample stock returns of Table 1. On Figure 2a the optimal depth 2 RT split when the stock return vector is regressed on f2 twice is shown (i.e. the RT is restricted to use only f2 as a predictor at both levels). As before, the 1st split is based on f2 < 4.5, and then in internal nodes #1 and #4, split criteria f2 < 1.4 and f2 < 5.4 are respectively used. Residual MSE at depth 2 for the RT equals ⅜ · 1.167 + ⅜ · .722 = .70833. Similarly, on Figure 2b the optimal depth 2 RT split when the stock return vector is regressed on f1 twice is shown. In nodes #1 and #4, f1 < 2.1 and f1 < 5.5 are the splitting criteria. Residual MSE for the RT equals ⅜ · 1.056 + ¼ · .562 + ¼ · .25 = .60. On Figure 2c the optimal depth 2 RT split when the stock return vector is freely regressed on both factors (i.e. f1, and f2) is shown. Initially, factor f1 provides more information so the split on the 1st level is based as before on f1<=4.9. Then, in nodes #1 and #4, f2 is used.

*Figure 2a*. Split for the optimal depth 2 RT when the target vector is regressed on f2 twice (only f2 at both levels).

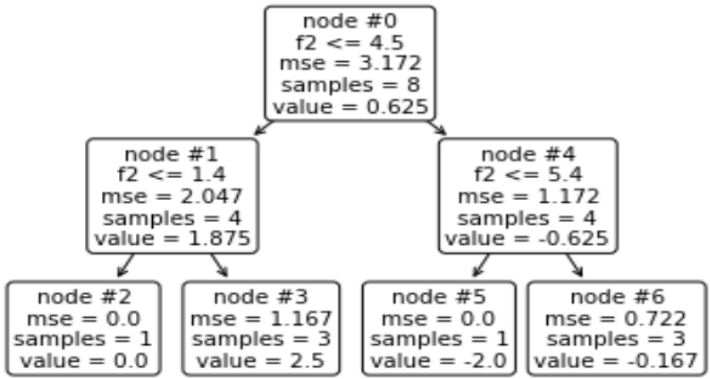

Residual MSE for the RT equals ¼ · .562 + ⅜ · .389 + ¼ · .25 = .348875 which is better than .71 for f2 and .6 for f1, and even below the benchmark .92 1-bit RT with y regressed on itself (i.e., 1-bit lepto-variance). The 2-bit lepto-variance is calculated on Figure 2d via the optimal 2-bit benchmark RT when y is the predictor. Using y to sort, both extreme target values y = -2 and y = 4 are isolated. Residual MSE for the RT equals $\lambda 2^2$ = ⅜ · .167 + ⅜ · .167 = .125 which is lower than the .35 residual MSE for f1+f2 combined. The 2-bit macro-variance of y equals $\mu 2^2$=3.172 - .125 = 3.047.

*Figure 2b*. Split for the optimal depth=2 RT when the stock return vector is regressed on f1 twice (only f1 at both levels).

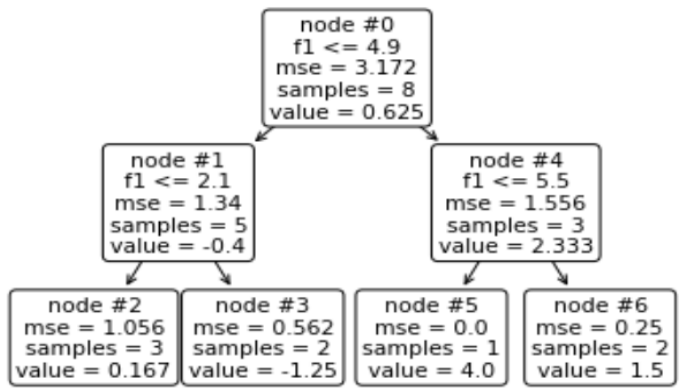

*Figure 2c. Split for the optimal depth=2 RT when the stock return vector is freely regressed on both factors (i.e., f1, and f2).*

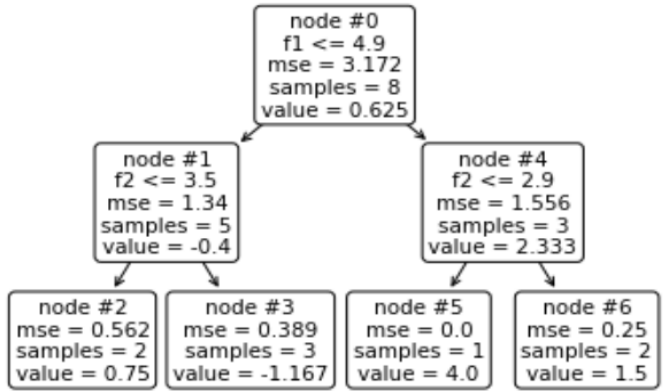

*Figure 2d. Splits for the benchmark depth=2 RT when the stock return vector is twice regressed on itself. Residual MSE is the 2-bit lepto-variance of y.*

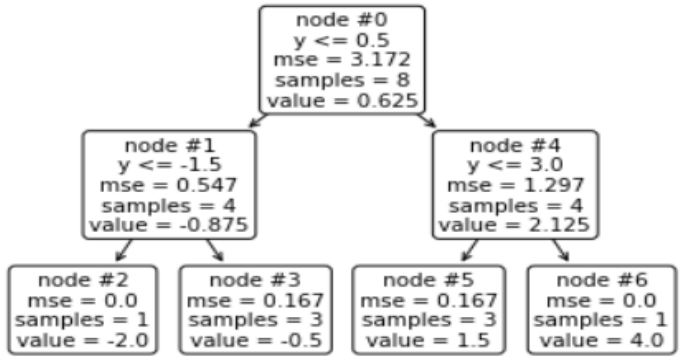

## 3. EXAMPLE LEPTO-VARIANCE ESTIMATION FOR A STOCK

The concepts of lepto- and macro-variance may be of interest for a host of machine learning applications from biostatistics to genomics. Here, to further demonstrate the concept of lepto-variance an example from financial analysis, a 3-factor analysis with real stock return data, is performed. Specifically, 5 years of daily return data for a major US corporation, International Business Machines Corp. (IBM), for the period from 1/5/2015 to 30/4/2020 are analyzed. The sample comprises 1259 daily returns.

In financial asset pricing, a well-known set of features is the three-factor model of [Fama and French (1993)]. The three-factor model expands the simpler Capital Asset Pricing Model by [Sharpe (1964)] and [Lintner (1965)] based on a time-series linear regression of excess portfolio returns of the type

$$y(t) = a + b \cdot MEx(t) + s \cdot SMB(t) + h \cdot HML(t) + e(t) \qquad (2)$$

y(t) is the excess return on a security for period t above the risk-free return. Factor MEx is the excess return on the value-weight (VW) market portfolio above the risk-free asset. Factor SMB is the return on a diversified portfolio of small stocks minus the return on a diversified portfolio of big stocks for the period. Factor HML is the difference between the returns on diversified portfolios of high and low book-to-market (B/M) ratio stocks. The three-factor linear model above assumes that the sensitivities b, s and h in (2) capture most variation in expected returns, so the true value of the intercept in (2) should equal the risk-free rate for all well-priced securities or portfolios.

Basic descriptive statistics for the daily returns of the 3 Fama-French factors and IBM stock for this 5-year period are shown in Table 1b. We utilize the lepto-variance framework developed here to demonstrate the relative power of the Fama and French three factors in explaining IBM return variability. Table 1c presents the correlation matrix of the three factors and IBM.

*Table 1b. Descriptive statistics for the daily returns of the 3 Fama-French factors and IBM stock for the 5-year period from 1/5/2015 to 30/4/2020 (in percentages)*

|       | MEx        | SMB        | HML        | IBM        |
|-------|------------|------------|------------|------------|
| count | 1259.000000 | 1259.000000 | 1259.000000 | 1259.000000 |
| mean  | 0.036704   | -0.017752  | -0.034662  | 0.004972   |
| std   | 1.201973   | 0.588960   | 0.666216   | 1.563978   |
| min   | -12.000000 | -4.580000  | -4.710000  | -12.850000 |
| 25%   | -0.325000  | -0.340000  | -0.390000  | -0.615000  |
| 50%   | 0.050000   | -0.040000  | -0.060000  | 0.050000   |
| 75%   | 0.520000   | 0.305000   | 0.300000   | 0.665000   |
| max   | 9.340000   | 5.730000   | 3.190000   | 11.300000  |

*Table 1c. Correlation matrix of the 3 factors and IBM returns.*

|     | MEx      | SMB      | HML      | IBM      |
|-----|----------|----------|----------|----------|
| MEx | 1.000000 | 0.145484 | 0.137149 | 0.734029 |
| SMB | 0.145484 | 1.000000 | 0.218836 | 0.045696 |
| HML | 0.137149 | 0.218836 | 1.000000 | 0.147349 |
| IBM | 0.734029 | 0.045696 | 0.147349 | 1.000000 |

The 1-bit lepto-structure of IBM is recovered via a depth 1 RT of IBM returns on IBM itself. The optimal depth 1 RT when IBM is regressed on itself is shown in Figure 3a. At the root of the tree (node #0) the entire sample is included with mean return .005, and total MSE of 2.444. Optimal split is for IBM < -.365 which is valid for 32.6% of the sample. The remaining 67.4% of the samples belong to the right child. Remaining depth 1 lepto-variance equals $\lambda 1^2$ = .326 x 1.894 + .674 x 1.234 = 1.449. The 1-bit **macro-variance** of IBM equals the MSE drop from the original 2.444 total sample MSE, $\mu 1^2 = 2.444 – 1.449 = .995$.

Next we run a depth 1 RT where the IBM return vector is regressed on all 3 Fama-French factors (i.e., the market excess return MEx and the SMB and HML factors). The optimal depth 1 RT is shown on Figure 3b. The MEx factor dominates with a residual MSE = .137 x 2.984 + .863 x 1.771 = 1.937. Thus, MEx explains 2.444 - 1.937= .507 of IBM variance which represents .507/.995 = 51% of the total IBM 1-bit macro-variance.

*Figure 3a. The optimal depth 1 RT when the IBM return vector (IBM) is regressed on itself.*

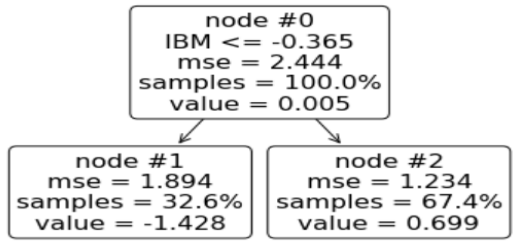

*Figure 3b. Optimal depth 1 RT when the IBM return vector is regressed on all 3 factors*

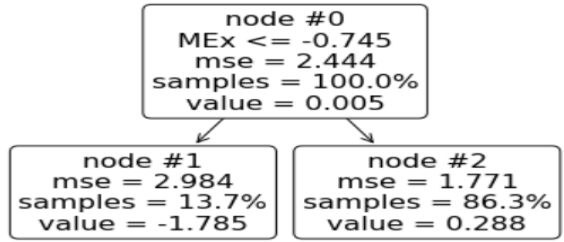

*Figure 3c. Optimal depth 1 RT for IBM regressed only on the Fama-French SMB and HML factors.*

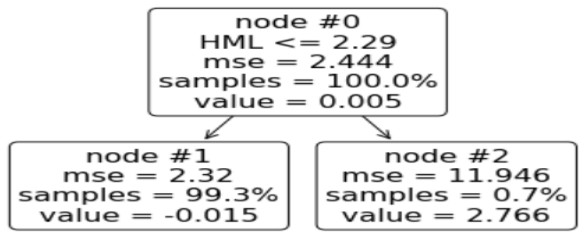

Next, we run a depth 1 RT where the IBM return vector is regressed on the Fama-French SMB and HML factors (i.e., the MEx feature is excluded from the RT to force the use of SMB or HML). The optimal depth 1 RT is shown on Figure 3c. Between the 2 features, HML dominates SMB and has a residual MSE of .993 x 2.32 + .007 x 11.946 = 2.389. Thus, HML explains .055 of the total MSE which represents .055/.995 = 5.5% of IBMs 1-bit macro-structure. Finally, we run a depth 1 RT where the IBM return vector is regressed only on the SMB factor (i.e., both MEx and HML features are excluded from the RT to force the use of SMB). The optimal depth 1 RT is shown on Figure 3d. Use of the SMB factor generates a residual MSE of .033 x 11.224 + .967 x 2.117 = 2.413. Thus, SMB only explains 3.12% of the 1-bit macro-variance for IBM.

**Figure 3d**. *Optimal depth 1 RT when the IBM return vector is regressed only on the SMB factor*

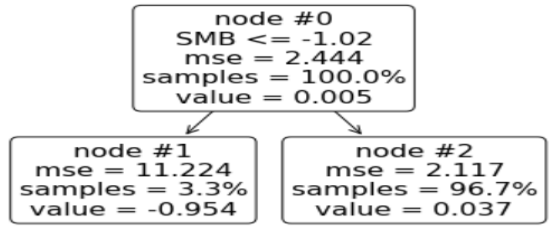

## 3.1 The 2-bit Lepto-structure for IBM

We now focus on the 2-bit lepto-variance for IBM. On Figure 4a, the split for the optimal depth 2 RT when the IBM return vector is regressed on itself is shown. The 2-bit lepto-variance is the residual MSE $\lambda_2^2$ = .04 x 3.328 + .287 x .295 + .628 x .339 + .045 x 3.642 = .594. This means that a 2.444 - .594 = 1.85 has been explained using a 2-bit RT. The first bit explained $\mu_1^2$ = .995, and the 2nd level explained an extra $\mu_2^2$ - $\mu_1^2$ = 1.85 - .995 = .855. In total, 75.7% = 1.85/2.444 of total IBM variability is macro-structure at the 2-bits level, while the remainder 24.3% is the 2-bit lepto-structure.

**Figure 4a**. *Split for the optimal depth 2 RT when the IBM return vector is regressed on itself.*

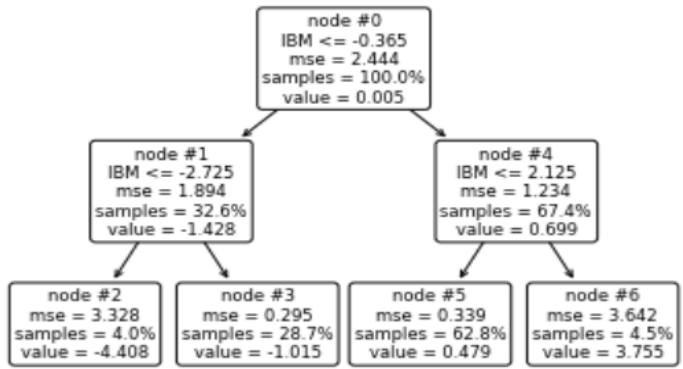

On Figure 4b, the split for the optimal depth 2 RT when the IBM return vector is regressed on the 3 factors is shown. The Market return factor dominates with a depth 2 residual MSE = 1.466, implying a total explained variance .978/1.85 = 52.86% of the full 2-bit IBM macro-structure.

On Figure 4c, the split for the optimal depth 2 RT when the IBM return vector is regressed on the Fama-French SMB and HML factors is shown. Residual variance equals 2.325, implying a total explained variance at level 2 of only .119/1.85 = 6.43% of the 2-bit macro-variance. Finally, the optimal split (not shown) for the 2-bit RT when the IBM return vector is regressed only on the SMB factor (the lowest correlation feature) has residual MSE = 2.380, implying a total explained variance at level 2 of only .064/1.85 = 3.46% of the 2-bit macro-variance.

**Figure 4b**. Split for the optimal depth 2 RT when the IBM return vector is regressed on the 3 factors.

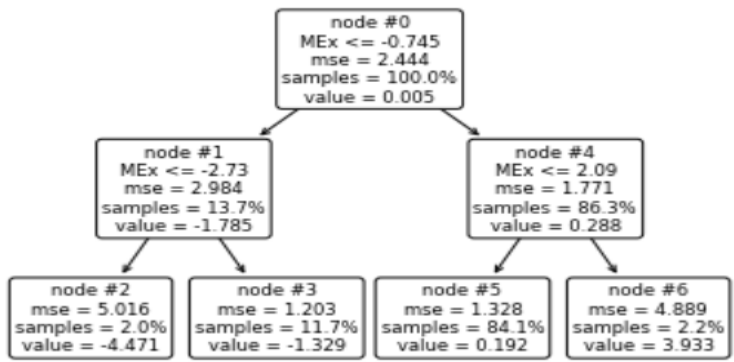

**Figure 4c**. Split for the optimal depth 2 RT when the IBM return vector is regressed on the Fama-French SMB and HML factors.

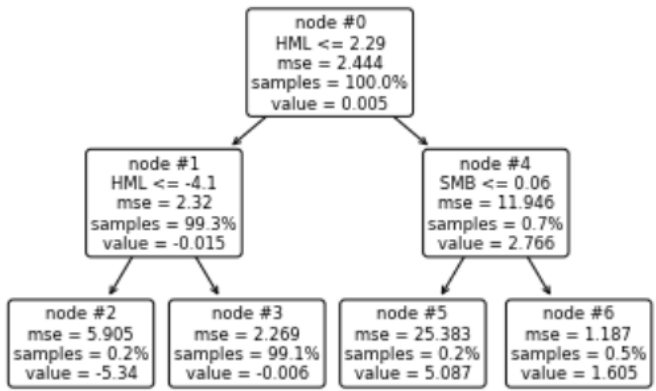

# 4. CONCLUSION

It is shown here that since in a Regression Tree (RT) it is always beneficial to generate a sorted split of a sample S, using the target itself as predictor provides an upper bound in terms of the sample variability that can be explained. The k-bit lepto-variance of a sample is defined as the residual MSE after the sample has been regressed on itself up to k times. The k-bit macro-variance is the variance captured by the target itself, and thus represents the maximum variance that can be captured by any combination of features. Total sample variance is decomposed in macro- and lepto-variability. The existence of lepto-variance in a sample places a limit on the resolving power of a k-bit RT in explaining its fine structure. As a demonstration, the lepto-structure analysis of 5-years of daily IBM stock returns is performed. It is shown that while market movements may explain slightly more than 50% of the return macro-structure, the SMB and HML factors can only explain negligible amounts of the 1-bit and 2-bit macro-structure of IBM returns.

---

[i] The term variance in this paper implies a population variance (simple sum of squared error division by sample size n) and not the variance estimate which corrects for degrees of freedom.

[ii] when the split is after position 8 all samples belong in the same child and the MSE is the total variance for the sample 3.172 (exactly 3.171875).

[iii] This conjecture seems intuitive but is not obvious and needs to be formally shown.